\documentclass[aps,twocolumn,floats,prl,nofootinbib,superscriptaddress,10pt]{revtex4-1}

\usepackage[utf8]{inputenc}

\usepackage[dvips]{graphicx} %
\usepackage{graphicx,amsmath,amsfonts,amssymb,slashed}
\usepackage{bbold,wasysym}
\usepackage{graphicx}

\usepackage[usenames,dvipsnames]{xcolor}

\usepackage{soul}
\usepackage{float}
\usepackage{amssymb, amsmath, epsfig}
\usepackage{graphics,times,array,ctable,amsmath,amssymb,natbib}
\usepackage{threeparttable}
\newcommand{\HeII}{He{\sc \, ii}}
\newcommand{\HeIII}{He{\sc \, iii}}

\def\apj{ApJ}
\def\apjl{ApJL}

\def\mnras{MNRAS}

\def\araa{{Ann.\ Rev.\ Astron.\& Astrophys.\ }}

\def\prd{PRD}

\begin{document}

\title{Nucleosynthetic signatures of primordial origin around supermassive black holes}

\author{Phoebe Upton Sanderbeck}
 \email{phoebeu@lanl.gov}
 \affiliation{Los Alamos National Laboratory, Los Alamos, NM}
\affiliation{Department of Physics \& Astronomy, University of California, Riverside}

\author{Simeon Bird}%

\affiliation{Department of Physics \& Astronomy, University of California, Riverside}

\author{Zolt\'{a}n Haiman}

\affiliation{Department of Astronomy, Columbia University, New York, NY}

\date{\today}

\begin{abstract}
  If primordial black holes (PBHs) seeded the supermassive black holes (SMBHs) at the centers of high-redshift quasars, then the gas surrounding these black holes may reveal  nucleosynthetic clues to their primordial origins. We present predictions of altered primordial abundances around PBHs massive enough to seed SMBHs at $z\approx 6-7.5$. We find that if PBHs with initial masses of $\sim 10^5$ M$_{\odot}$ are responsible for such SMBHs, they may produce primordial Deuterium and Helium fractions enhanced by $\geq 10\%$, and Lithium abundance depleted by $\geq 10\%$, at distances of up to $\approx$ a comoving kiloparsec away from the black hole after decoupling. We estimate that $\sim$ 10$^8$M$_{\odot}$ of gas is enhanced (or depleted) by at least one percent.
 Evidence of these modified primordial Deuterium, Helium, and Lithium abundances could still be present if this circum-PBH gas remains unaccreted by the SMBH and in or near the host galaxies of high-redshift quasars.  Measuring the abundance anomalies will be challenging, but could offer a novel way to reveal the primordial origin of such SMBH seeds.  
  
\end{abstract}

\maketitle

\section{Introduction}

The number of quasars observed above $z=6$ has exceeded $200$, with several discovered at $z>7$ \cite{inayoshi19,bosman20}
including a $z = 7.5$ quasar with a $1.5\times 10^{9}$M$_{\odot}$ black hole \cite{yang20}. The mechanism by which the supermassive black holes (SMBHs) driving these
quasars are assembled within a Gyr of the Big Bang is poorly understood. The simple picture of Eddington-limited accretion of massive Pop III seed
black holes cannot reasonably produce the SMBHs seen at high redshift, so
super-Eddington growth, an increase in the mass of the black hole seeds, and growth via frequent mergers have been proposed as potential solutions to this tension \cite{inayoshi19}.

Invoking primordial black holes (PBHs) as seeds for these high-redshift SMBHs can plausibly solve the problem of early SMBH assembly, as PBHs as large as $10^5 {\rm M}_{\odot}$ that form relatively late can plausibly make up a small fraction of the dark matter \cite{bean02,rice17,carr18}. The mass of a PBH forming at a time $t$ would have a mass that roughly matches the particle horizon size that goes as $\sim10^5 (t/s) {\rm M}_{\odot}$-- a population of PBHs that form around $t \sim$1 second would have formed from the collapse of the steepest density peaks, leaving these objects sufficiently massive \cite{carr18}. The number density of these late-forming PBHs depends on the perturbation amplitude on these very small scales.

If any of the SMBHs hosted by $z>6$ quasars are indeed seeded by PBHs, we
have to consider that there exists a modulation to the local expansion rate around these PBHs. This would lead to inhomogeneous Big Bang Nucleosynthesis (BBN), where these rare, high-density peaks would induce inhomogeneities in the Helium fraction, as well as other primordial abundances \cite{carr18}. Such an inhomogeneity was examined recently in \cite{arbey20}, who found that signatures of overdense regions at BBN could lead to unusual abundance patterns in early stellar populations. Here we investigate observable signatures of inhomogeneous BBN around massive PBHs, assumed to be potential SMBH seeds.

We show that if SMBHs observed above $z\approx$ 6 are seeded by rare massive PBHs, the primordial abundances in proximity to
the SMBH will be modulated by the PBH's density perturbation at BBN. Consequently, we may be able to constrain the history of SMBH formation and the possibility of PBHs seeding SMBHs by observationally probing the chemical abundances around SMBHs. This requires that the circum-PBH  gas with anomalous abundances ``survives'' -- i.e. that it is not rapidly accreted by the SMBH and that it remains in the SMBH's vicinity at lower redshift.

\section{Big Bang Nucleosynthesis}

\begin{figure}
\resizebox{9.5cm}{!}{\includegraphics{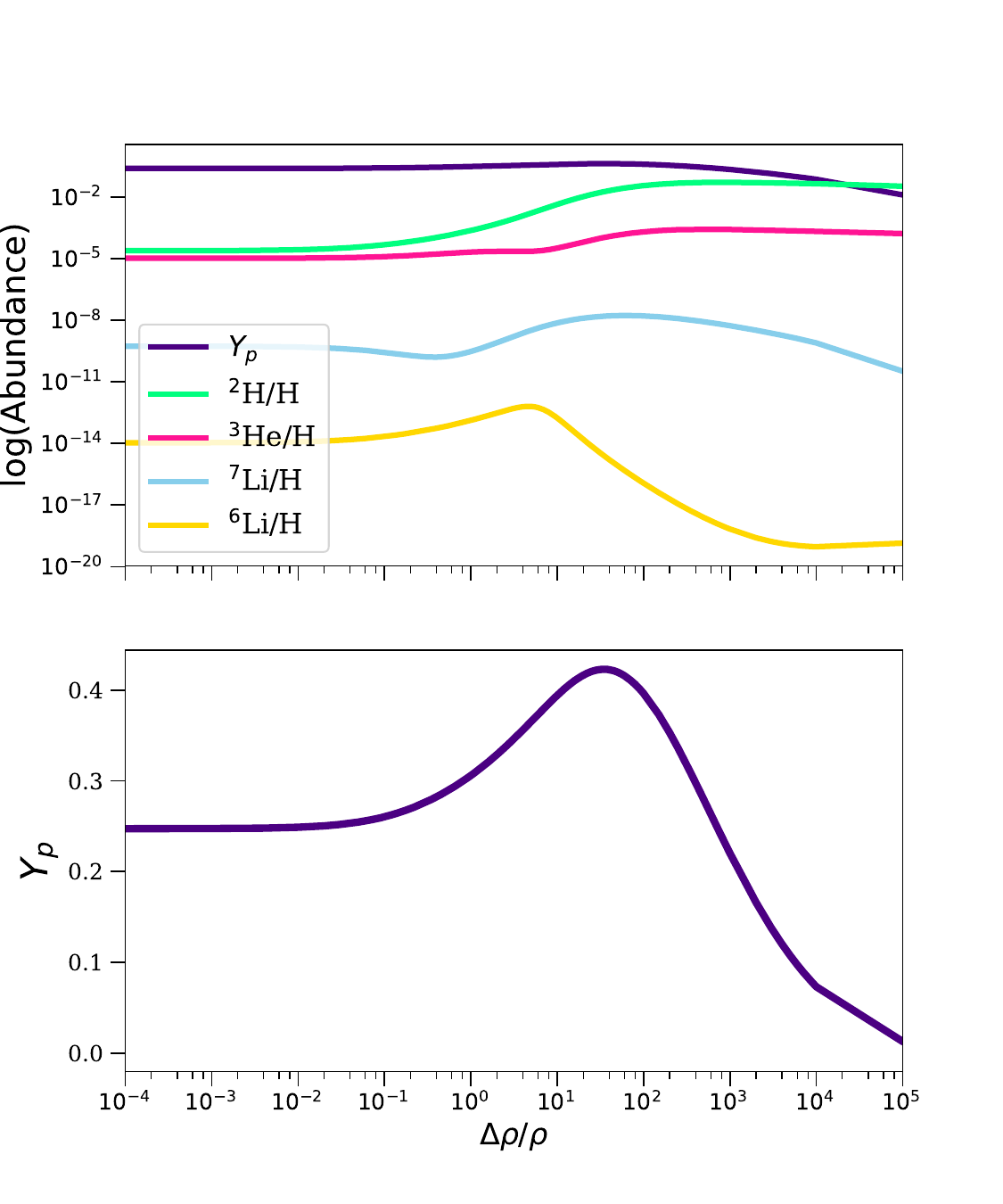}}\\
\caption{Primordial abundances as a function of density in excess of the critical density as calculated from {\tt AlterBBN}. The top panel shows the relative abundances of all primordial elements through Lithium. The lower panel zooms in on the Helium abundance.
\label{fig:Yp}}
\end{figure}

\begin{figure*}
\begin{center}
\resizebox{19.5cm}{!}{\includegraphics{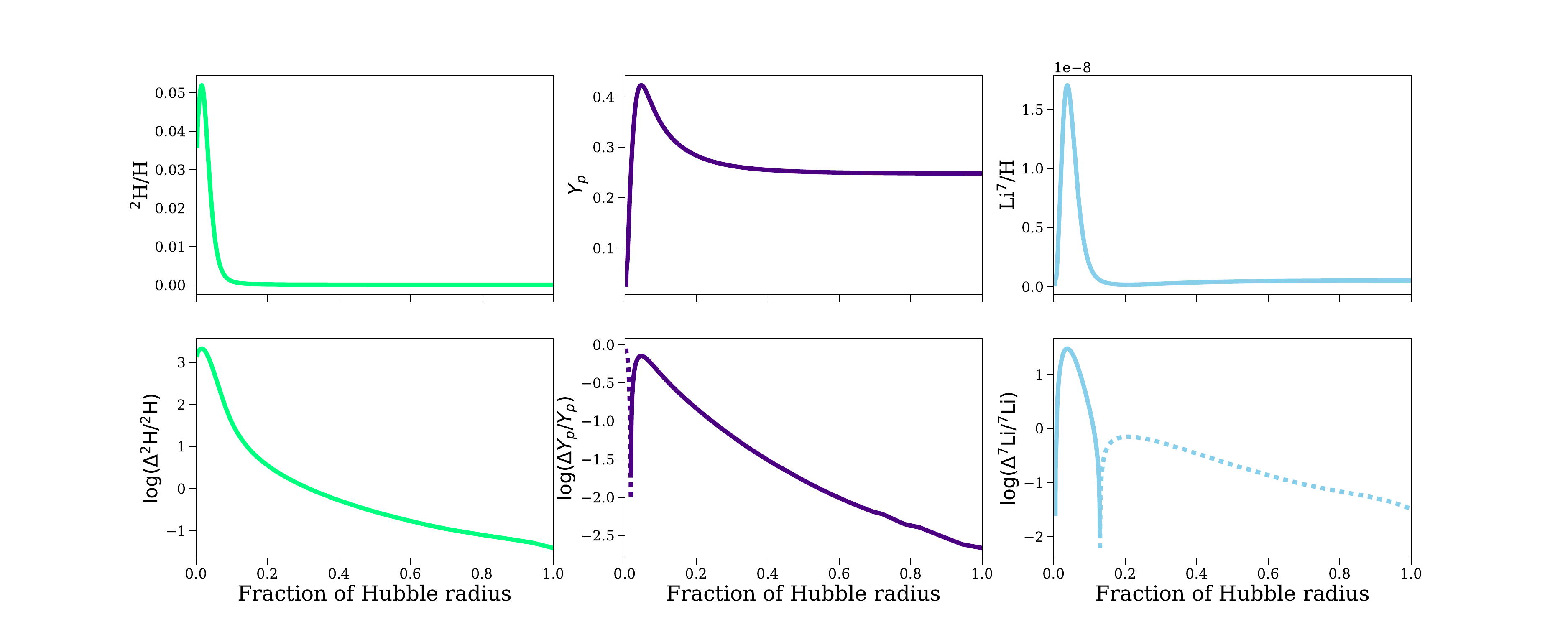}}\\
\end{center}
\caption{\ {\bf Upper left panel:} the primordial Deuterium abundance by unit mass, $^2$H/H, at BBN as a function of distance from a $10^5$ M$_{\odot}$ PBH. The Hubble radius at BBN is $\approx 0.6$ AU. {\bf Lower left panel:} the log of the fractional change in $^2$H/H as seen in the upper panel. {\bf Upper middle panel:} the primordial Helium abundance by unit mass, $Y_p$, at BBN in the same region. {\bf Lower middle panel:} the log of the fractional change in $Y_p$ as seen in the top panel. {\bf Upper right panel:} The primordial Lithium abundance in the same region.
  {\bf Lower right panel:} the log of the fractional change in $^7$Li/H as seen in the top panel.  The unperturbed abundance of Deuterium is $^2$H/H $= 2.435\times 10 ^{-5}$, for Helium is $Y_p = 0.2473$, and for Lithium-7 is $^7$Li/H $= 5.467\times10^{-10}$. In the lower panels, dotted lines represent depletion, and solid lines an enhancement in the elemental abundance.
\label{fig:BBN_distr}}
\end{figure*}

The presence of a PBH at BBN will perturb the local expansion rate such that
\begin{equation}
\label{eq:expansion}
H^2  = \frac{8\pi }{3} \left[\rho + \Delta\rho\right],
\end{equation}
where $\rho$ is the critical density, dominated at this time by radiation, and $\Delta \rho$ is the local density enhancement from a proximate PBH, which we treat as a point potential so that the local densities of dark matter and baryons are not altered. A modulated expansion rate at BBN will induce inhomogeneities in primordial abundances that we investigate as follows.

We use the publicly available Big Bang Nucleosynthesis code {\tt AlterBBN} \cite{abbn18,abbn12} to calculate primordial chemical abundances. {\tt AlterBBN} allows users to consider non-standard cosmological scenarios, including modifications to the expansion rate. Figure~\ref{fig:Yp} shows the results of these calculations when we introduce an additional density component, $\Delta\rho$, in excess of the critical density.  The top
panel shows the relative abundances of all primordial elements through Lithium and the bottom panel focuses only on the Helium fraction, $Y_p\equiv {\rm He}/({\rm H}+{\rm He})$. As the local density increases, the Helium fraction grows until an overdensity of $\Delta\rho/\rho\approx 35$, where an increase in Deuterium causes the Helium fraction to shrink.

\begin{figure}
\resizebox{9.0cm}{!}{\includegraphics{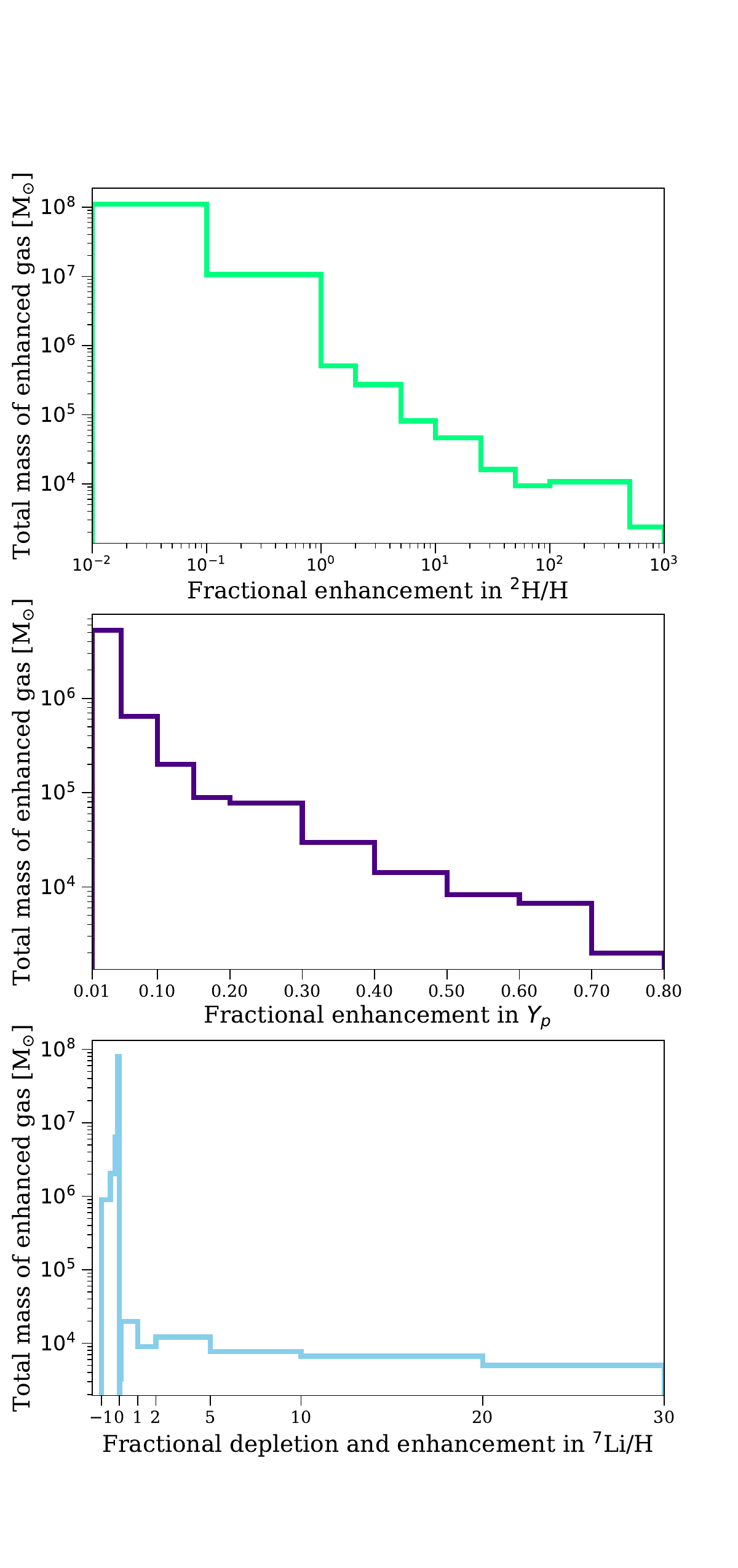}}\\
\caption{{\bf Upper panel:} the distribution of Deuterium-enhancement
fraction in the primoridal gas at BBN around our 10$^5$M$_{\odot}$ PBH. Each bin shows the mass of the gas enhanced to levels shown on the $x$ axis.
{\bf Middle panel:} the census of Helium-enhanced gas formed at BBN around our PBH. For clarification, the first bin represents gas that has a 1-5\% percent enhancement above the standard primordial Deuterium abundance.
  {\bf Lower panel:} the census of $^7$Li-depleted and enhanced gas. Note
that most of the gas is $^7$Li-depleted rather than enhanced. The fine bins in the lower panel represent fraction depletions of 1 to 0.5, 0.5 to 0.25, 0.25 to 0.1, and 0.1 to 0.01; and fractional enhancements of 0.01 to 0.1. Note that we have omitted all gas that is enhanced or depleted by less than one percent.
\label{fig:masses}}
\end{figure}

\begin{figure*}
\begin{center}
\resizebox{19.5cm}{!}{\includegraphics{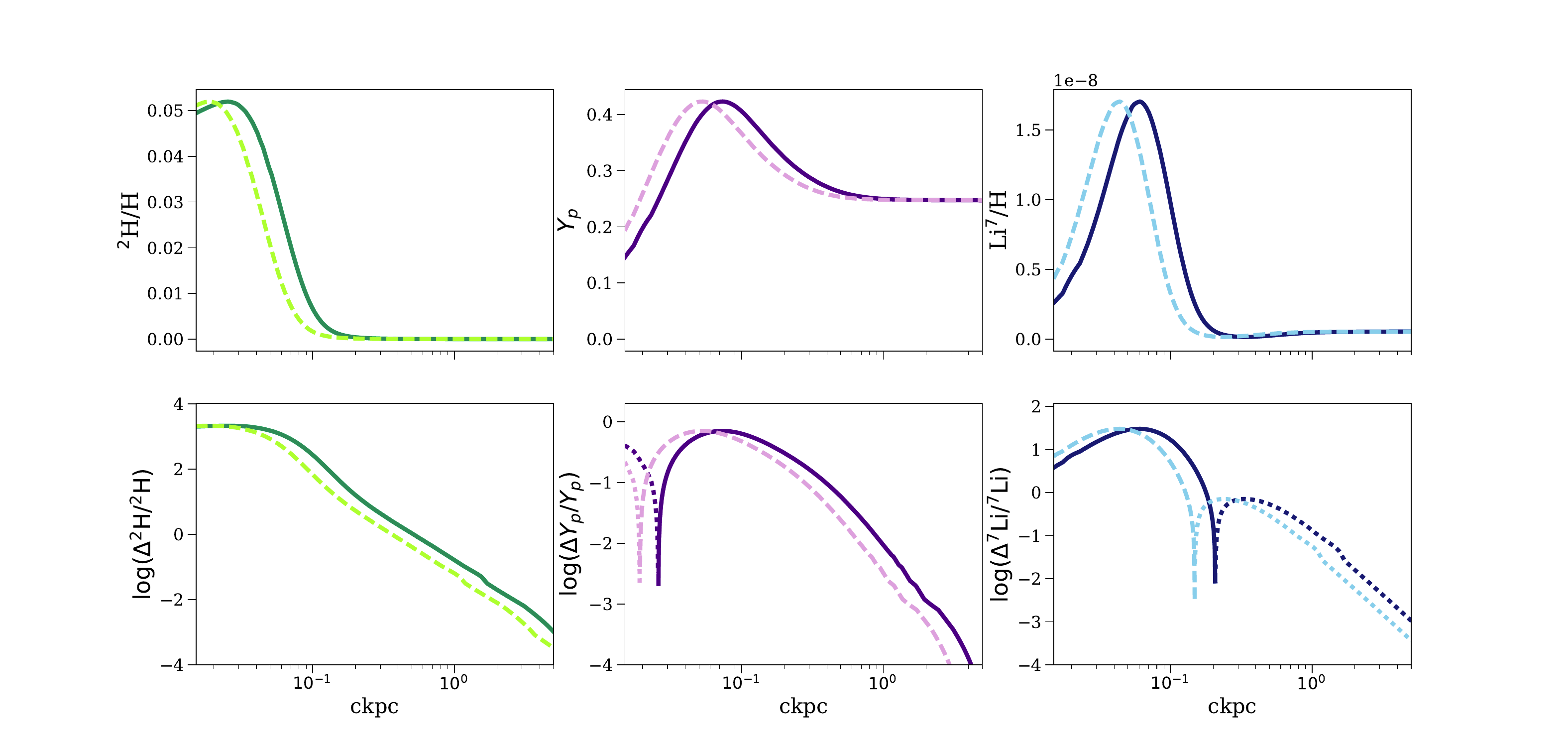}}\\
\end{center}
\caption{The upper row shows the spatial distributions of light primordial element abundances at $z=1100$ (decoupling) as a function of distance from the PBH, including the mass fractions for Deuterium $^2$H/H ({\bf left}),  Helium $Y_p$ ({\bf middle}) and Lithium $^7$Li/H ({\bf right}). The bottom row shows the corresponding (logarithm) of the fractional changes relative to the global average abundance of each species.
The unperturbed background abundaces are $^2$H/H $= 2.435\times 10 ^{-5}$, $Y_p = 0.2473$ and $^7$Li/H $= 5.467\times10^{-10}$. The solid curves show our model with an initial density perturbation. The baryons and radiation are coupled up until this point, pushing the baryons away from the initial point perturbation at the sound speed -- roughly $c/\sqrt{3}$ during radiation domination. The dashed curves show, for reference, a model where we have assumed that the gas distribution simply follows the Hubble flow with time. The dotted portion of the curves in the lower panels represent depletions rather than enhancements in Helium and Lithium.
\label{fig:decoupling}}
\end{figure*}

Using the abundance trends in Figure~\ref{fig:Yp}, we then estimate the change in these abundances around a PBH at BBN. As previously motivated, we set our PBH mass to $10^5{\rm M}_{\odot}$, the birth mass of a PBH formed at $t\sim 1$ second. We then take the fractional change in total density surrounding the PBH relative to the critical density as a function of a fraction of the Hubble volume, and interpolate this fractional change in density with our abundance trends shown in Figure~\ref{fig:Yp}.

Figure~\ref{fig:BBN_distr} shows our predictions for Deuterium ($^2$H), Helium ($^4$He), and Lithium ($^7$Li) abundances at BBN ($t\sim 3$ minutes) surrounding a 10$^5$ M$_{\odot}$ PBH. The top left panel of Figure~\ref{fig:BBN_distr} shows the primordial Deuterium abundance by unit mass, $^2$H/H, around our $10^5{\rm M}_{\odot}$ PBH at BBN within approximately a Hubble radius (note, at BBN the Hubble radius, $r_{H}=c/H(z)$, is approximately 0.6 AU).
The lower left panel shows the same, but presented as the log of the fractional change of $^2$H/H from its unperturbed value, to highlight the abundance gradient at larger radii. The PBH induces a $\gtrsim$ factor of two change in the Deuterium abundance out to $\sim50\%$ of a Hubble radius away from the PBH, and over an order of magnitude change out to $\sim 20$\% of the Hubble radius.

The top middle panel of Figure~\ref{fig:BBN_distr} shows the primordial Helium abundance by unit mass, $Y_p$, around our $10^5{\rm M}_{\odot}$ PBH
at BBN within approximately a Hubble radius. The lower middle panel shows the log of the fractional change in $Y_p$. The PBH induces a $\gtrsim $ one percent change in the Helium fraction out to more than half of a
Hubble radius away from the PBH, with a maximum change in $Y_p$ at $\sim 10$ percent of the Hubble radius from its location, and $\sim 20$ percent increase in $Y_p$ out to $\sim 20$ percent of the Hubble radius.

Finally, the top right in Figure~\ref{fig:BBN_distr} shows the primordial Lithium-7 fraction, $^7$Li/H, around our $10^5{\rm M}_{\odot}$ PBH at BBN where the unperturbed Lithium-7 abundance is $^7$Li/H = $5.467\times10^{-10}$. The PBH induces a maximum change in $^7$Li/H at $\sim 5$ percent of the Hubble radius from its center, with considerable modulation out to the Hubble radius. The lower right panel shows the log of the fractional change in $^7$Li/H. We find a 10\% modulation in Lithium abundance out to approximately a Hubble radius from the PBH, with a maximum change in $^7$Li/H of $30\times$ at $\lesssim 10$ percent of the Hubble radius.

Although we have shown how these primordial abundances change with distance from the PBH, it is useful to take stock of how much gas falls within different abundance bins. To estimate the density profile of gas surrounding the black hole at BBN, we solve simultaneously for the perturbations in cold dark matter, baryons, and photons, using the tight-coupling approximation \cite{peebles70} when the Thomson opacity is sufficiently large.\footnote{Specifically, when $\tau_c/\tau\ll 1$, where $\tau_c = an_e \sigma_T$ and $\dot{a}/a \sim 1/\tau$. } We subsequently Fourier transform the solutions
to obtain the appropriate density profiles in position space. We model the black hole as part of the dark matter with a delta function density profile. This assumes the black hole is non-interacting, which is a good approximation at this redshift \cite{alih17}. Assuming that everything within the horizon collapses into the PBH when the PBH forms, we set the initial conditions such that the baryons and radiation are unperturbed from their mean cosmic values. 

Figure~\ref{fig:masses} shows the breakdown of mass in each of our selected abundance bins for Helium (middle panel), Deuterium (top panel), and Lithium-7 (lower panel). We show all gas with a Deuterium abundance enhanced by more than one percent, where we find approximately $10^8$M$_{\odot}$ of such enhanced gas. The middle panel shows Helium gas enhanced by more than one percent, with approximately $10^{6.7}$M$_{\odot}$ of total gas above this threshold. In the lower panel, we show both Lithium-7-depleted and enhanced gas, where approximately $10^{8}$M$_{\odot}$ of gas is depleted by more than one percent, and approximately $10^{4.7}$M$_{\odot}$ of gas is enhanced by more than ten percent. The fine bins in the lower panel in Figure~\ref{fig:masses} represent fraction depletions of 1 to 0.5, 0.5 to 0.25, 0.25 to 0.1, and 0.1 to 0.01; and fractional enhancements of 0.01 to 0.1. We have omitted all gas that is enhanced or depleted by less than one percent.

\section{Decoupling}

While Figure~\ref{fig:BBN_distr} shows the scale of the abundance inhomogeneities at BBN, we must extend these predictions to lower redshift, where it may be possible to observationally detect such an effect. Prior to decoupling, the baryons couple to the radiation and act as a single fluid. The photon-baryon interactions in an overdensity (from an initial perturbation) overpressurize the fluid, resulting in a spherical acoustic wave that propagates outward from the perturbation. This pushes the baryons away from the initial point perturbation at the sound
speed (approximately $c/\sqrt{3}$ during radiation domination). As gas travels away from the initial perturbation, carried by the spherical acoustic wave,
this initial region becomes underdense in gas and the black hole and dark matter overdensity do not attract gas back until well after decoupling.

Up until the epoch of decoupling, the PBH should accrete minimally due to the overpressured photon-baryon plasma. Thus we make the approximation that all circum-PBH gas at BBN survives accretion until decoupling. After decoupling, if we approximate the PBH's accretion rate by the spherical Bondi rate, $\dot{M} = \pi \rho G^2 M^2 / c_s^3$, we find that a $10^5$ M$_{\odot}$ PBH should accrete on the order of 0.1 M$_{\odot}$yr$^{-1}$ at $z\approx 1100$ (where $c_s\approx5~{\rm km~s^{-1}}$), and $5\times 10^{-3}~{\rm M}_{\odot}$yr$^{-1}$ at $z\approx 10$ ($c_s\approx0.15~{\rm km~s^{-1}}$). At this rate, a $10^5$ M$_{\odot}$ PBH could roughly accrete 20 times its own mass, mostly at low redshifts, which is a small fraction ($\lesssim 10\%$) of the $10^{7-8}~{\rm M_\odot}$ of modified-abundance circum-PBH material. Furthermore, this represents an upper limit, because it ignores several mechanisms that suppress accretion, such as radiation or wind feedback, nonzero angular momentum of the gas, or non-spherical symmetry. Additionally, prior to decoupling, PBHs will have some velocity relative to the gas due to the matter-radiation coupling that induces supersonic relative velocities between dark matter and baryons. This will effectively further suppress accretion onto PBHs until after decoupling~\cite{alih17}.  

Although the non-linearity of low-redshift ($z\lesssim 1000$) evolution becomes more complicated than can be described through cosmological perturbation theory, we numerically solve the coupled linearized Einstein, Boltzmann, and fluid equations to describe the spatial evolution of the gas around the black hole from BBN until decoupling. As described previously, we solve simultaneously for the perturbations in cold dark matter, baryons, and photons in the Newtonian gauge in Fourier space (we have neglected neutrinos). During the tight coupling regime, we solve the following coupled differential equations found in \cite{callin06} (Equations 22, 33, and 34)\footnote{Also see Section 5.7 in \cite{ma95}} using a numerical ODE solver,

\begin{eqnarray}
\label{eq:expansion}
\dot{\delta}_{\rm CDM} = \frac{k}{\mathcal{H}}v-3\dot{\phi} \nonumber \\
\dot{v}_{\rm CDM} = -v_{\rm CDM}-\frac{k}{\mathcal{H}}\dot{\psi} \nonumber\\
\dot{\delta}_{b} = \frac{k}{\mathcal{H}}v_{b} - 3\dot{\phi} \nonumber\\
\dot{v}_{b} = -v_{b} - \frac{k}{\mathcal{H}}\psi +\dot{\tau} R \left(3\theta_{1} + v_{b}\right) \nonumber\\
\dot{\phi} = \psi - \frac{k^2}{3\mathcal{H}^2}\phi + \frac{H_0^2}{2\mathcal{H}^2}\left(\frac{\Omega_{m}\delta_{\rm CDM}}{a} + \frac{\Omega_{b}\delta_{b}}{a} +\frac{4\Omega_{r}\theta_0}{a^2}\right) \nonumber\\
 \psi = -\phi - \frac{12 H_0^2}{k^2a^2}\Omega_r \theta_{2} \nonumber\\
\dot{\theta}_{0} = -\frac{k}{\mathcal{H}}\theta_{1} - \dot{\phi} \nonumber\\
\dot{\theta}_{1} = \frac{1}{3}\left(\left[3\dot{\theta}_{1}+\dot{v}_{b}\right]-\dot{v}_{b}\right), \nonumber\\ 
\end{eqnarray}
where $\delta$ and $v$ are the density perturbation and velocity shown for cold dark matter and baryons (denoted by their subscripts), $\theta_{N=0,1,2}$ are the photon temperature multipoles, $\phi$ and $\psi$ are the scalar potentials, and $\tau$ is the optical depth as a function of redshift. Here we define $\mathcal{H}=\dot{a}/a$, and $R =4\Omega_{r}/(3\Omega_{b})$.

The tight coupling approximation allows us to substitute

\begin{eqnarray}
\left[3\dot{\theta}_{1}+\dot{v}_{b}\right] = \nonumber \\
\frac{-\left[(1-R)(\dot{\tau}+\ddot{\tau})\right](3\theta_{1}+v_{b})-\frac{k}{\mathcal{H}}\psi}{ (1+R)\dot{\tau}+\frac{\dot{\mathcal{H}}}{\mathcal{H}}-1} \nonumber \\
+\frac{\left(1-\frac{\dot{\mathcal{H}}}{\mathcal{H}}\right)\frac{k}{\mathcal{H}}(-\theta_{0}+2\theta_{2}) +\frac{k}{\mathcal{H}}(-\dot{\theta}_0 +2\dot{\theta}_2)}{ (1+R)\dot{\tau}+\frac{\dot{\mathcal{H}}}{\mathcal{H}}-1}
 \end{eqnarray}
 
 and 
 \begin{equation}
 \theta_{2} =- \frac{20k}{45\mathcal{H}\dot{\tau}}\theta_{1}.
\end{equation}

After the tight-coupling regime, we include higher order photon temperature multipoles for $l\geq 2$,

 \begin{equation}
 \dot{\theta}_l = \frac{lk}{(2l+1)\mathcal{H}}\theta_{l-1}-\frac{(l+1)k}{(2l+1)\mathcal{H}}\theta_{l+1} +\dot{\tau}\left(\theta_l-\frac{1}{10}\theta_2\delta_{l,2}\right),
 \end{equation}
up to $l=4$. Finally, we Fourier transform the solutions of these coupled equations to obtain the appropriate density profiles in position space.


To track the spatial evolution of a particular gas parcel at a given radius, we assume spherical symmetry, and invoke a simple conservation of mass argument. We define the mass within a given radius from the black hole at a given redshift as
\begin{equation}
M_{\rm gas}(r,z) = \int_{0}^r \left[1+\delta_{b}(r,z)\right] \rho_{b}(z) 4\pi r^2 dr 
\end{equation}
where $\delta_b$ is the baryonic density perturbation, $\rho_{b}=\Omega_b\rho_{c}$, and $\rho_{c}$ is the critical density at redshift $z$. We then interpolate between the primordial abundances as a function of radius, $^2$H/H$(r)$, $Y_p(r)$, and $^7$Li/H$(r)$, as they are shown in Figure~\ref{fig:BBN_distr}, and $M_{\rm gas}(r,z_{\rm BBN})$. This relation then allows us to
extrapolate the spatial fluctuations of enhanced abundances to lower redshifts, where we assume that the $M_{\rm gas}-^2$H/H, $M_{\rm gas}-Y_p$,
and $M_{\rm gas}-^7$Li/H relations hold until structure formation becomes nonlinear, shells cross, and gas mixes substantially between scales.

Figure~\ref{fig:decoupling} shows the primordial Deuterium (left panels),
Helium (middle panels), and Lithium (right panels) abundances around our PBH at decoupling ($z\approx 1100$). The upper panels show $^2$H/H$(r)$, $Y_p(r)$, and $^7$Li/H$(r)$, while the lower panels show the log of the fractional change in these abundances. The gas surrounding the PBH with modulated $^2$H/H, $Y_p$, and $^7$Li/H has been carried out to approximately a comoving kpc away by the acoustic wave of coupled photons and baryons, leaving an underdense region surrounding the central black hole. After decoupling, the baryons stall, as the photon pressure dissipates due to lack of Thomson scattering. Thus, at decoupling, the affected gas will likely have reached its maximum (comoving) radius away from the central PBH.  The dashed curves show a scenario where we have assumed that the gas distribution simply follows the Hubble flow with time.

Following decoupling, the spatial evolution of this gas becomes more uncertain. Structure formation becomes nonlinear and the PBH is likely to start accreting appreciably $\sim100$ Myr later. By $z=10$, the gravitational instability induced by the perturbation has attracted new baryons and dark matter to the overdensity \cite{eisenstein07}, rendering our conservation of mass argument invalid. The largest spatial scale of the effect shown in Figure~\ref{fig:decoupling} ($\sim $ ckpc) indicates that the modified gas is likely within the host galaxy that would have recently formed around the PBH. The total mass of enhanced gas is $\sim 10^8$M$_{\odot}$, which is one percent of the total mass of a $10^{10}$ M$_{\odot}$ SMBH, roughly the maximum mass of a SMBH found in quasars found at $z>6$.

The final distribution of modified abundance gas around SMBHs in quasars at $z\approx 6-7$ hinges on several unknown factors. First, it is dependent on the morphology of large-scale structures at the location of the quasar host. 
Second, the fate of this gas will be
dependent on the details of formation of the quasar host. When the black hole begins to substantially gain mass during matter domination, the mode
of this accretion could determine how much of the enhanced gas survives to low redshift. A 10$^5$M$_{\odot}$ PBH that begins to accrete at the Eddington limited rate at $z\approx30$ can grow into 10$^{10}$M$_{\odot}$ SMBH by $z=7.5$ assuming a radiative efficiency $\lesssim 10\%$, which may swallow most of the enriched gas. However, it is likely that SMBHs grow by both mergers and accretion: merger-driven growth and/or aspherical accretion
could lead to some of the enriched gas remaining extant around the SMBH.

\section{Observability}

Should large and rare PBHs seed some of the SMBHs found at $z>6$, and should enhanced gas avoid total accretion onto the black hole prior to observation, the detectability of the resulting fluctuations in primordial abundances would provide compelling evidence for large and rare PBHs as
the seeds of these high-redshift SMBHs. There are a number of avenues to potentially look for enhanced and depleted primordial Deuterium, Helium, and Lithium. For instance, if the gas stays local to the host quasar/galaxy as indicated by Figure~\ref{fig:decoupling}, investigation of the stellar population in the host galaxy or the accretion disk. However, due to the likely limited quantity of surviving enhanced gas, it will be difficult to make this detection.

Any remnant gas with anomalous abundances will likely reside entirely within a bright quasar's host galaxy.  This could lead to systematically stronger broad Helium emission lines (eg. He~1640 and other HeII recombination lines) \cite{oh01,yang06}. Future narrow-band observations with JWST could be used to detect such lines~\cite{oh01}. Because the strength of this line is dependent on nonlinear astrophysics, a small modulation in the primordial Helium fraction would need to be disentangled from other effects, such as Helium lines powered by the hard spectra of Population III stars~\cite{tumlinson00}. However, a sufficiently large sample of quasars may statistically reveal an increased line strength for the highest redshift and most massive quasars. The Helium-enhanced gas would be most observable with
the He1640 line if it were to fall within the broad-line region, close to
the SMBH, but prior to accretion.  It would be distinct from narrower stellar He lines. Additionally, other Galactic sources such as metal-poor halo stars, stars in old globular clusters, and dwarf galaxies have already been identified as plausible targets for studying modified primordial abundances~\cite{arbey20}.

In principle, the hyperfine transition of $^3$\HeII\ at 8.66 GHz (3.5 cm) could be detected with future radio surveys and, in conjunction with HI 21 cm observations, could probe the primordial Helium abundance \cite{bagla09,khullar20}.  Although this signal is weak, future radio surveys with sufficient survey volume and wavenumber bandwidth could detect it during the epoch of reionization, when \HeII\ is abundant~Ref.~\cite{khullar20}.  In practice, in any surviving primordial gas near the quasar BH, Helium is likely to be doubly ionized (\HeIII), unless it is shielded from the quasar's far-UV radiation, or the quasar is exceptionally young, making this detection unlikely.
Measuring fluctuations of primordial Lithium appears equally challenging, as typically primordial Lithium abundances are derived from metal-poor Galactic stars.   While neutral Lithium could impact the cosmic microwave background~\cite{loeb01}, ultraviolet photons from the recombination of hydrogen keep Lithium ionized when it would have otherwise recombined around $z\sim 300-500$ \cite{switzer05}, making this effect unmeasurable in practice.


While the anomalous Deuterium, Helium, and Lithium abundances appear challenging to measure, detecting these fluctuations could provide compelling evidence for large PBHs as seeds of the SMBHs found at $z\sim 6-7.5$.   



\begin{acknowledgments}

We thank the anonymous referee for their helpful comments and suggestions. P.U.S. and S.B. were supported by NSF grant AST-1817256 and Z.H. was supported by NSF grant AST-2006176. We would like to thank Ryan Cooke for helpful discussions. P. U. S. would also like to acknowledge support of a LDRD Director’s Postdoctoral Fellowship at Los Alamos National Laboratory.

\end{acknowledgments}

\bibliography{PBH}


\end{document}